\documentclass{ws-p9-75x6-50}

\begin{document}

\title{Formation of optical lines in black-hole binaries }

\author{Kinwah Wu,}
\address{
  School of Physics, University of Sydney, NSW 2006, Australia and \\ 
  Mullard Space Science Laboratory, University College London, \\ 
  Holmbury St.~Mary, Dorking, Surrey RH5 6NT, United Kingdom, \\ 
  E-mail: kw@mssl.ucl.ac.uk}

\author{Roberto Soria,}
\address{Mullard Space Science Laboratory, 
  University College London, \\ Holmbury St.~Mary, Dorking,  
  Surrey RH5 6NT, United Kingdom, \\ 
  E-mail: rs1@mssl.ucl.ac.uk} 

\author{Helen Johnston and Richard Hunstead}
\address{School of Physics, University of Sydney, NSW 2006, 
  Australia \\ 
  E-mail: helenj@physics.usyd.edu.au, rwh@physics.usyd.edu.au }  
 
\maketitle

\abstracts{ 
The H~{\scriptsize I} Balmer emission lines of black-hole binaries 
  show double-peaked profiles during the high-soft state 
  and the quiescent state. 
In the high-soft state the profiles are asymmetric 
  with a stronger red peak, 
  but the profiles are symmetric in the quiescent state. 
We suggest that in the high-soft state 
  the emission lines originate 
  from the temperature-inversion layer caused by 
  irradiative heating of an optically thick accretion disk. 
Irradiative heating also causes the formation of a disk wind, 
  which mildly absorbs the blue peak of the lines. 
The double-peaked lines seen in the quiescent state 
  arise from an optically thin disk. 
In the absence of a disk wind, the lines are unabsorbed 
  and so the symmetry of the line profiles is preserved. }  

\section{Background}   

Many soft X-ray transients are probably close binary systems 
  containing a black hole accreting material 
  from a low-mass companion star.  
These systems often show three distinctive X-ray spectral states,  
  which are commonly classified as 
  the high-soft state, the low-hard state and the quiescent state 
  (see e.g. review by Tanaka and Lewin\cite{tlxx95}). 
The high-soft state is characterised  
  by a prominent blackbody-like thermal component 
  and high-energy power-law tail in the X-ray spectrum.    
The effective temperature of the thermal component is $\sim 1$~keV.   
The slope of the power-law is steep, 
  with a photon index $\Gamma \sim 2-4$, 
  and it can extend to energies beyond 100~keV.   
The thermal component is generally insignificant in the low-hard state.  
The X-ray spectrum is dominated by a power-law component, 
  but the photon index of the power-law (typically $\sim 1-2$) 
  is significantly flatter than that of the high-soft state.  

There is strong observational evidence 
  that the morphology of the optical emission lines 
  changes with the X-ray spectral properties\cite{sxxx01}. 
For instance, it was found that 
  the equivalent widths of H~{\scriptsize I} Balmer emission lines 
  correlated with the hard (20--100~keV) X-ray luminosity 
  during the hard X-ray outbursts of GRO~J1655$-$40\cite{swhx00}. 
This correlation was, however, not found 
  for the He~{\scriptsize II} lines\cite{swhx00}.     
Another example is that  
  the H~{\scriptsize I} Balmer emission lines of black-hole binaries 
  generally have double-peaked profiles   
  in the high-soft state\cite{swjx99}.  
The two peaks are asymmetric, 
  with the red peak stronger than the blue peak.   
It also appears that the double-peaked lines 
  are in superposition with a broader absorption trough\cite{swhx00}.   
This is contrary to the situation in the low-hard state, 
  when the H~{\scriptsize I} Balmer lines are more often single-peaked\cite{swjx99}.   
Interestingly, when the systems are in quiescence 
  the lines are also double-peaked (e.g.\ A0620$-$00), but 
unlike the asymmetric profiles seen in the high-soft state,  
  the line profiles tend to be very symmetric\cite{jkox89}.   

\section{Formation of double-peaked lines}   

A direct interpretation is that 
  the double-peaked lines arise from the accretion disk\cite{sxxx81}.  
The red and blue peaks come from 
  the disk regions 
  with receding and approaching line-of-sight velocities respectively.  
When the system is X-ray active,   
  it is also optically bright, 
  with the optical continuum dominated by the disk emission. 
The optical emission region in the accretion disk is probably opaque, 
  as the optical continuum seems to be well fitted 
  by a thermal blackbody model\cite{hhsc98}.    

The standard theories of radiative transfer 
  (see e.g.\ Mihalas\cite{mxxx78}, Tucker\cite{txxx76}) predict 
  a smooth thermal spectrum for the emission 
  from an opaque, isothermal medium. 
The spectrum would show absorption line features    
  if there is a negative temperature gradient towards the surface. 
Emission lines are formed 
  only when the brightness temperature at the line-centre energies 
  is higher than the thermal temperature of the nearby continuum.    
If the medium is opaque to the
   continuum, this implies that the lines are formed 
  in the presence of a temperature-inversion surface layer. 
This is analogous to the spectra of OB stars, 
  which are dominated by absorption lines, 
  and the spectra of cooler G and K stars, which show strong emission lines.  
When the medium is transparent to the continuum, 
  the spectrum will show emission lines, because  
  the line opacities are generally higher than continuum opacities.  

If we accept the view that 
  the accretion disks in black-hole binaries are opaque to optical emission 
  in the X-ray active state, 
  then the double-peaked optical lines seen in the high-soft state 
  suggest the presence of a temperature-inversion layer 
  on the accretion-disk surface. 
The disappearance of the double-peaked profiles in the low-hard state 
  may be due either to the fact 
  that the geometry of the accretion disk---and 
  hence its optical depth---is modified\cite{sxxx01}, 
  or that the line emission is dominated by a non-disk component.  
The double-peaked lines seen in the quiescent state   
  are probably emission from a low-density accretion disk 
  optically thin to the continuum.   
We now examine this scenario in more detail.   

\section{Line formation in an opaque accretion disk}   

\subsection{A plane-parallel model} 

We use a simple plane-parallel model\cite{wshj01} 
  to illustrate that a temperature-inversion layer can be set up 
  on the surface of an accretion disk under soft X-ray illumination.  
The model is a modification of that proposed by Milne\cite{mxxx26} 
  for irradiative heating of the atmosphere of a star by its companion. 
The geometry of the model is shown in Figure 1.  
The basic assumptions in the model are  
  (i) the emission region is semi-infinite and planar, 
  (ii) the accretion disk is in radiative equilibrium, and 
  (iii) the process is linear.  
In addition, we also assume that the cooling 
  is dominated by the optical continuum emission.   

\begin{figure}[t]  
\begin{center} 
  \epsfxsize=16pc 
  \epsfbox{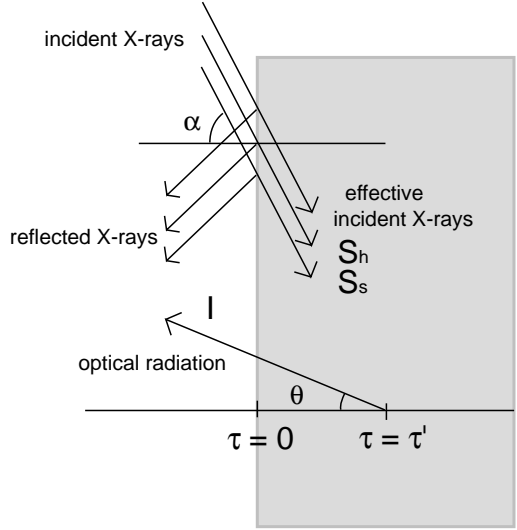}  
\end{center}  
\caption{The geometry of the plane-parallel model. 
  The disk surface is placed vertically and 
  is illuminated by soft and hard X-rays at an incident angle $\alpha$.   
\label{fig:plane}}    
\end{figure}    

As the radiative-transfer equations are linear  
  under these assumptions, 
  the temperature $T$ at an optical depth $\tau$ is given by 
\begin{eqnarray}  
    T(\tau) & = & \biggl[ {\pi \over \sigma} 
   \biggl( B_{\rm x}(\tau) + B_{\rm d}(\tau) \biggr) \biggr]^{1/4} \   
   \equiv \ \biggl[ {\pi \over \sigma} B(\tau) \biggr]^{1/4} \ ,  
\end{eqnarray}     
where $\sigma$ is the Stefan-Boltzmann constant, 
  $B_{\rm x}(\tau)$ is the component of the radiation 
  due to irradiative heating, and   
  $B_{\rm d}(\tau)$ is the component due to viscous heating 
  in the absence of X-ray irradiation.  
The disk component\cite{sxxx84} is  
\begin{equation}   
  B_{\rm d}(\tau)\ =\ {3  \over 4}S_{\rm d}~\biggl[ 
    \tau \biggl( 1 - {\tau \over {2 \tau_{\rm tot}}} \biggr) 
    + {2 \over 3} \biggr]   \ , 
\end{equation}    
  where $\tau_{\rm tot}$ is the total opacity of the accretion disk 
  in the direction perpendicular to the disk surface,  
  and $S_{\rm d} = \sigma T_{\rm eff}^4/ \pi$ 
  is the viscous energy flux from the disk mid-plane\cite{dlhc99}.  
The effective temperature $T_{\rm eff}$, which is determined  
  by balancing the energy generated by viscous heating and by 
  radiative loss, is given roughly by  
\begin{equation} 
  \sigma T_{\rm eff}^4 \  \approx \  
  {9 \over 8 } \nu \Sigma \Omega_{\rm K}^2  \ ,  
\end{equation}    
  where $\nu$ is the viscosity coefficient, 
  $\Sigma$ is the surface density of the disk, and 
  $\Omega_{\rm K}$ is the Keplerian angular velocity.  

Suppose the incident radiation consists of 
  parallel beams of soft and hard X-rays, 
  with effective fluxes $\pi S_{\rm s}$ and $\pi S_{\rm h}$ 
  per unit area normal to the beams, 
  and making an angle $\alpha$ with the normal to the disk plane. 
The beams are absorbed exponentially, 
  with the absorption coefficient of the soft X-rays $k_{\rm s}\kappa$, 
  and that of the hard X-rays $k_{\rm h}\kappa$, 
  where $\kappa$ is the absorption coefficient of the optical radiation.  
Radiative equilibrium requires 
  the rate of energy absorption 
  be equal to the rate of radiative cooling. 
It implies that  
\begin{equation} 
  \pi \big( k_{\rm s}S_{\rm s} e^{-k_{\rm s} \tau \sec \alpha} +  
   k_{\rm h}S_{\rm h} e^{-k_{\rm h} \tau \sec \alpha}\big) + \int_{4 \pi} 
   d\Omega~ I(\tau,\mu)\ 
   =\ 4\pi B_{\rm x}(\tau) \ ,  
\end{equation}  
  where $I(\tau,\mu)$ is the intensity of the optical radiation 
  from the irradiatively heated disk at 
  the depth $\tau$ in the direction $\theta~(= \cos^{-1}\mu)$.  
On the other hand, by multiplying the radiative-transfer equation 
\begin{equation} 
  \mu {d \over {d \tau}} I(\tau,\mu) \ 
   = \ I(\tau,\mu) - B_{\rm x}(\tau)   
\end{equation}  
  by a differential solid angle $d\Omega$, 
  integrating with respect to $\tau$ 
  and using the condition of radiative equilibrium, 
  we obtain  
\begin{equation} 
 \int_{4 \pi} 
   d\Omega~ \mu~I(\tau,\mu)  = \ 
 \pi \cos \alpha ~\big(
  {S_{\rm s}} ~ e^{-k_{\rm s} \tau \sec \alpha} +  
  {S_{\rm h}} ~ e^{-k_{\rm h} \tau \sec \alpha}\big) \ . 
\end{equation}    

For the condition of no inward component of $I(\tau,\mu)$ 
  at the boundary ($\tau = 0$),  
\begin{equation}  
  \int_{-1}^{0} d\mu~I(0,\mu) \ = \ 0     
\end{equation}   
and 
\begin{equation}  
  \int_{-1}^{0} 
   d\mu~\mu~I(0,\mu) \ = \ 0  \ .   
\end{equation}  
If we define a hardness parameter $\xi \equiv S_{\rm h}/S_{\rm s}$ and 
  a total X-ray flux $S_{\rm x} \equiv S_{\rm s} + S_{\rm h}$,  
  then by solving all the equations above 
  with the two boundary conditions, 
  we have 
\begin{eqnarray}  
 B_{\rm x}(\tau) \ = \ {1\over 2} S_{\rm x} ~\bigg\{ 
     k_{\rm s}  f_{\rm s}(\alpha) \biggl({\xi \over {1+ \xi}} \biggr) 
      \biggl[ \biggl({{\cos \alpha} \over {k_{\rm s}}}\biggr) 
     - \biggl({{\cos \alpha} \over {k_{\rm s}}} - {1\over 2} \biggr) 
       ~e^{- k_{\rm s}\tau \sec \alpha } \biggr]  &  & \nonumber \\ 
     + k_{\rm h} f_{\rm h}(\alpha)  \biggl({1 \over {1+ \xi}} \biggr) 
      \biggl[ \biggl({{\cos \alpha} \over {k_{\rm h}}}\biggr) 
     - \biggl({{\cos \alpha} \over {k_{\rm h}}} - {1\over 2} \biggr) 
       ~e^{- k_{\rm h}\tau \sec \alpha } \biggr] \bigg\} \ . & & 
\end{eqnarray}   
  where the functions $f_{\rm s}(\alpha)$ and $f_{\rm s}(\alpha)$  
  are given by 
\begin{equation} 
 f_{\rm s,h}(\alpha) \ = \ \biggl[~ 
   1 - \biggl({{\cos \alpha} \over {k_{\rm s,h}}}\biggr) 
   + \biggl({{\cos \alpha} \over {k_{\rm s,h}}}\biggr) 
   \biggl({{\cos \alpha} \over {k_{\rm s,h}}} - {1\over 2} \biggr)  
     \ln (1 + k_{\rm s,h} \sec \alpha ) ~\biggr]^{-1}   \ . 
\end{equation}   

\subsection{Temperature-inversion layer}    

In Figure 2 we show the temperature as a function of the optical depth 
  for annuli of accretion disks illuminated by soft and hard X-rays 
  at an incident angle of 87$^\circ$.  
When the disk is illuminated by soft X-rays,  
  if the X-ray flux is larger than the viscous heat flux 
  emerging from the disk mid-plane, 
  a temperature-inversion layer can be set up.  
If the X-ray flux is weak, 
  the temperature gradient remains negative, 
  implying that irradiative heating is unimportant 
  and the temperature structure is determined 
  mainly by the transport of the energy 
  generated deep in the disk mid-plane by viscous heating. 
Moreover, the temperature inversion is stronger 
  for larger incident angles of the X-rays\cite{wshj01}.  
When the disk is illuminated by hard X-rays, 
  the disk always has a negative temperature gradient,  
  and no strong temperature inversion occurs.

\begin{figure}[t]
\epsfxsize=15pc 
\epsfbox{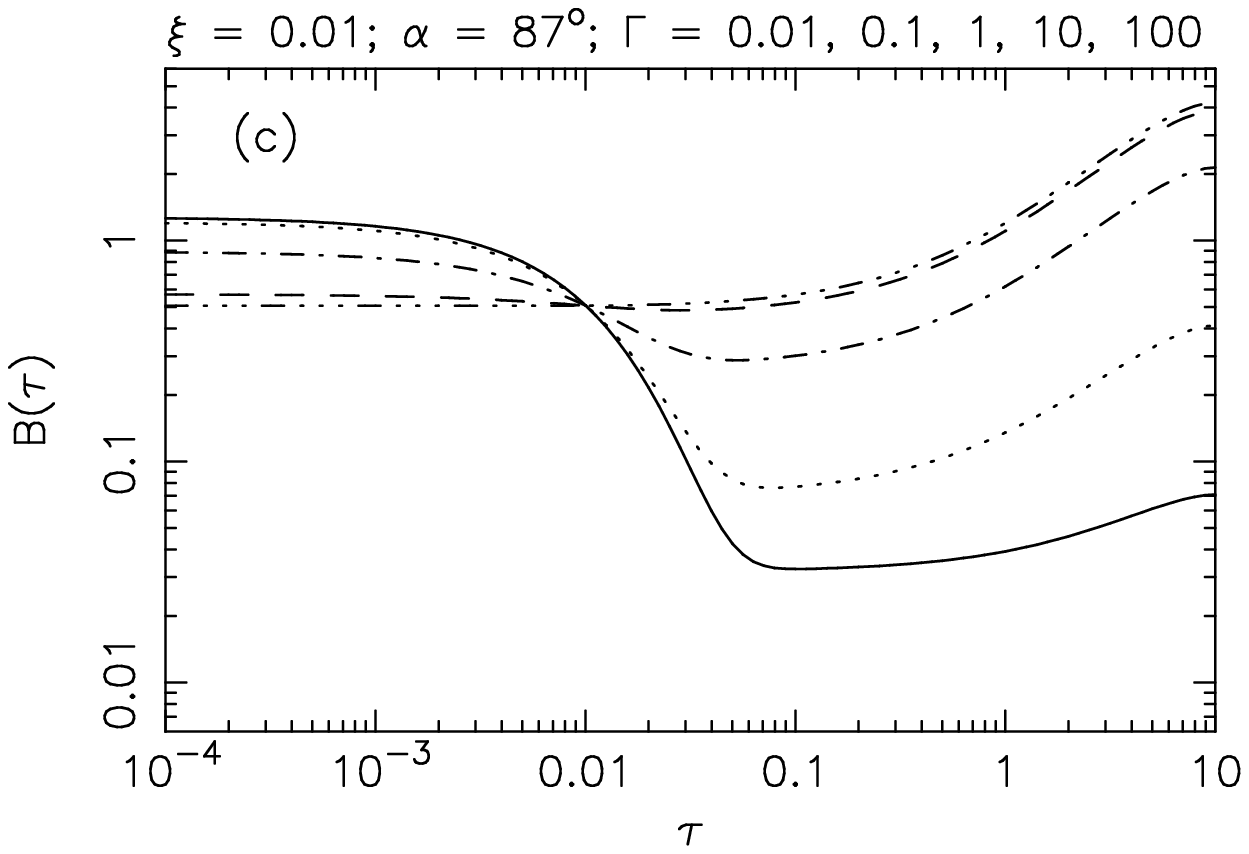}  
\epsfxsize=15pc 
\epsfbox{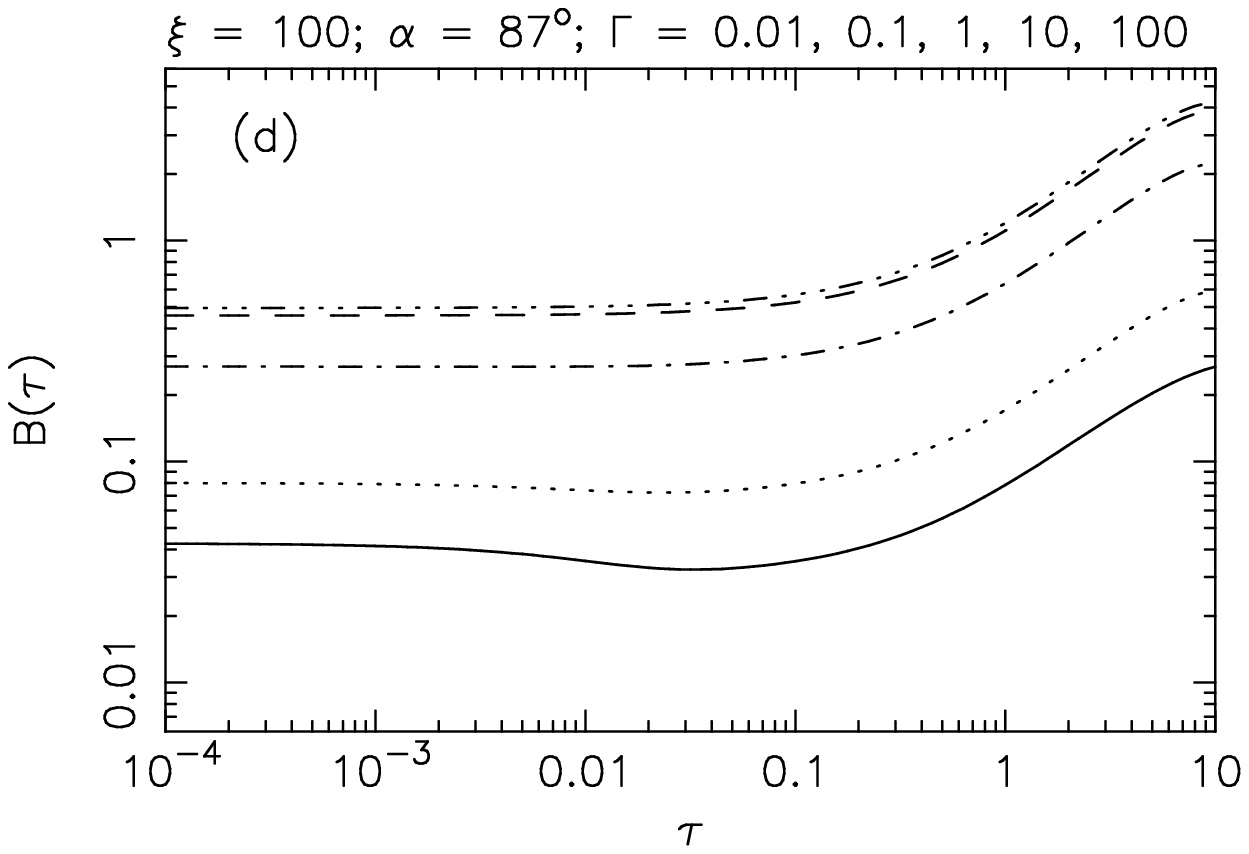}
\caption{
(Left) Temperature profile ($T \propto B^{1/4}$) 
  as a function of optical depth $\tau$ 
  for an X-ray incident angle $\alpha = 87^{\circ}$  
  and an hardness parameter $\xi = 0.01$ 
  (i.e.\ strong soft X-ray illumination).  
The two scaling parameters for the X-ray opacities are $k_{\rm s} = 5.0$ 
  and $k_{\rm h} = 0.01$, and  
  the total optical depth $\tau_{\rm tot} = 20$. 
The total flux density is normalised such that 
  $S_{\rm x} + S_{\rm d} = 1.0$. 
The curves correspond to 
  illumination-strength parameters  
  ${\tilde\Gamma} \equiv S_{\rm x}/S_{\rm d} = 100$  (solid line), 
  10 (dotted line), 1 (dash-dotted line), 
  0.1 (dashed line) and 0.01 (dash-dot-dot-dotted line).  
(Right) Same parameters as those in the left panel 
  except $\xi = 100$ (i.e.\ strong hard X-ray illumination). 
\label{fig:disk}}
\end{figure}     

Opacities are usually larger for soft X-rays ($\sim 0.1-1$~keV) 
  than for the optical continuum radiation, 
  and hence soft X-rays are more easily absorbed by the disk.   
When a significant amount of energy is deposited 
  at a small depth below the disk surface,   
  a temperature inversion is formed. 
Harder X-rays ($\sim 10$~keV), which are absorbed only in dense media, 
  tend to deposit energies deep below the disk surface. 
As a result, the disk is heated uniformly under hard X-ray illumination.  

In Figure 3 we show the location of disk regions 
  at which temperature inversion occurs 
  for systems with an accreting black hole of 7~M$_\odot$ 
  and a modest accretion rate of $5\times 10^{17}$~g~s$^{-1}$.  
The X-rays illuminate the accretion disk 
  at a grazing incident angle of 87$^\circ$   
  and a substantial portion of the X-rays are reflected.   
If the X-rays are hard (middle panel) and the reflection is large (99\%), 
  no strong temperature inversion can be set up 
  at all distances from the central black hole.  
The temperature structures in the inner disk regions ($R < 10^{10}$~cm)   
  are not significantly affected by irradiative heating.  
The temperatures of the outer disk regions ($R > 10^{10}$~cm), 
  however, increase substantially.   
For smaller reflection (50\%) and softer X-rays (right panel),  
  a weak temperature inversion is set up 
  at distances larger than $\sim 10^9$~cm from the black hole, and   
  the temperature structure of the disk region beyond $R \sim 10^8$cm   
  is greatly modified. 

If the X-rays are soft and the reflection is large, 
  a strong temperature inversion can be set up 
  over a substantial fraction of the area of the accretion disk 
  (left panel, Fig.~3).  
While the temperature structures of the inner disk regions  
  are still determined by viscous heating, 
  those of the outer disk rim are determined by irradiative heating. 
The predicted temperature of the temperature-inversion layer  
  at $R\sim 10^{11}$~cm is about $10^4-10^5$~K. 
This is indeed the temperature 
  at which the optical H and He lines are expected to form 
    (see e.g.\ Cox and Tucker\cite{ctxx69}, and Osterbrock\cite{oxxx89}).  
The distances predicted by the model ($R\sim 10^{11}$~cm) 
  are in good agreement with the observed peak-separation of GX~339$-$4 
  (left panel, Fig.~3)
  for an orbital period\cite{ccht92} of 15~h 
  and an orbital inclination\cite{wshj01} of 15$^\circ$.    
  
\begin{figure}[t]  
\begin{center} 
  \epsfxsize=9.5pc \epsfbox{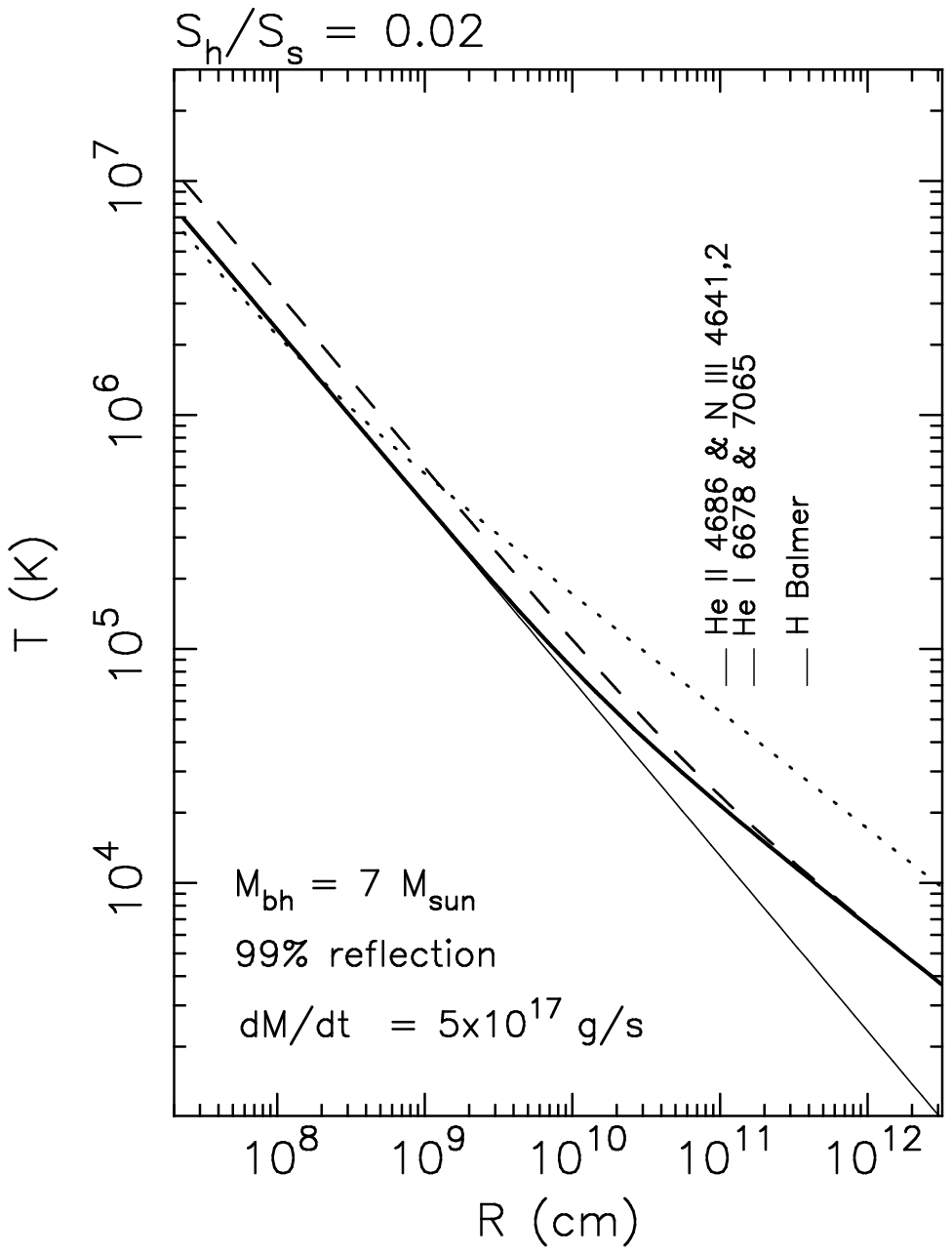} 
  \epsfxsize=9.5pc \epsfbox{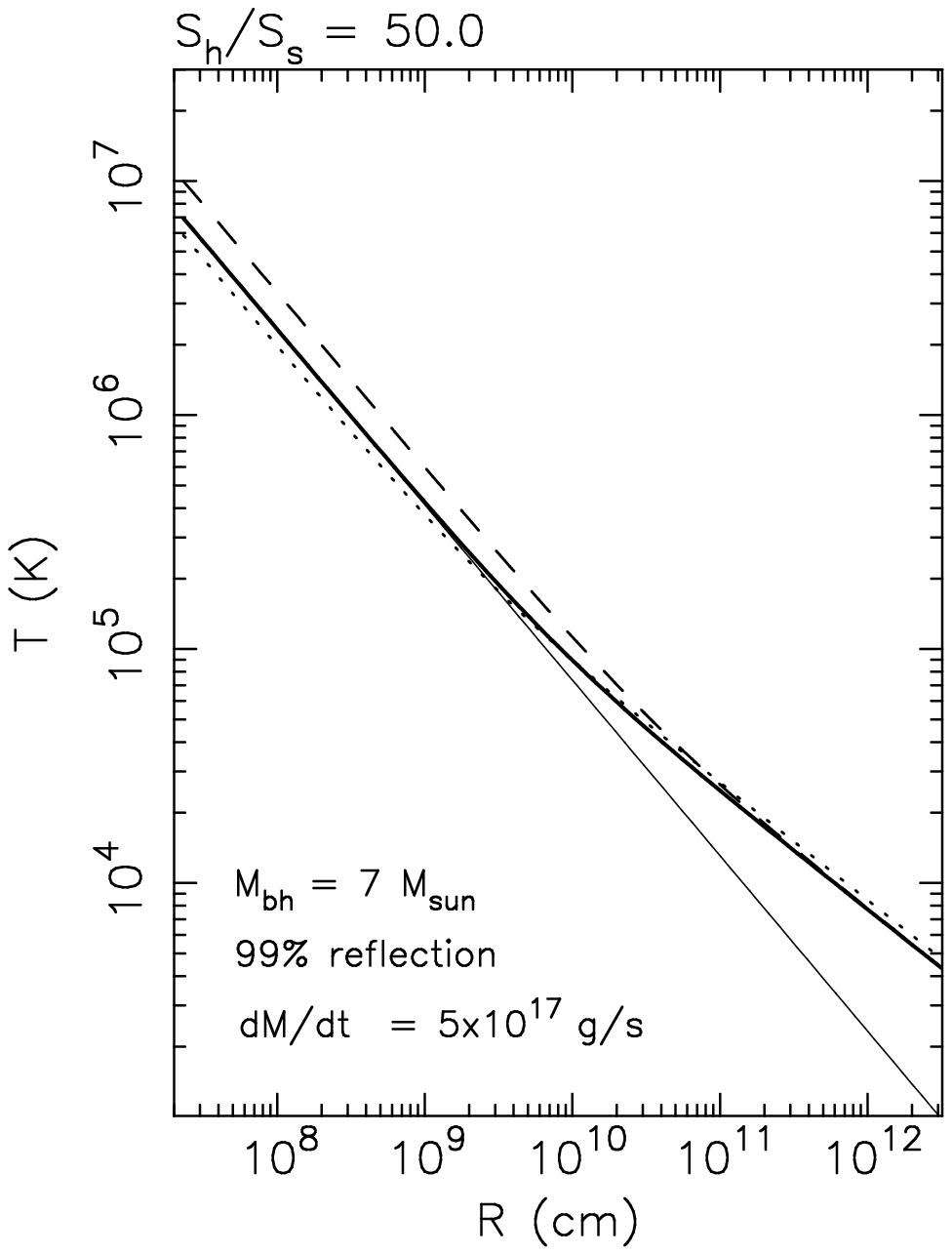} 
  \epsfxsize=9.5pc \epsfbox{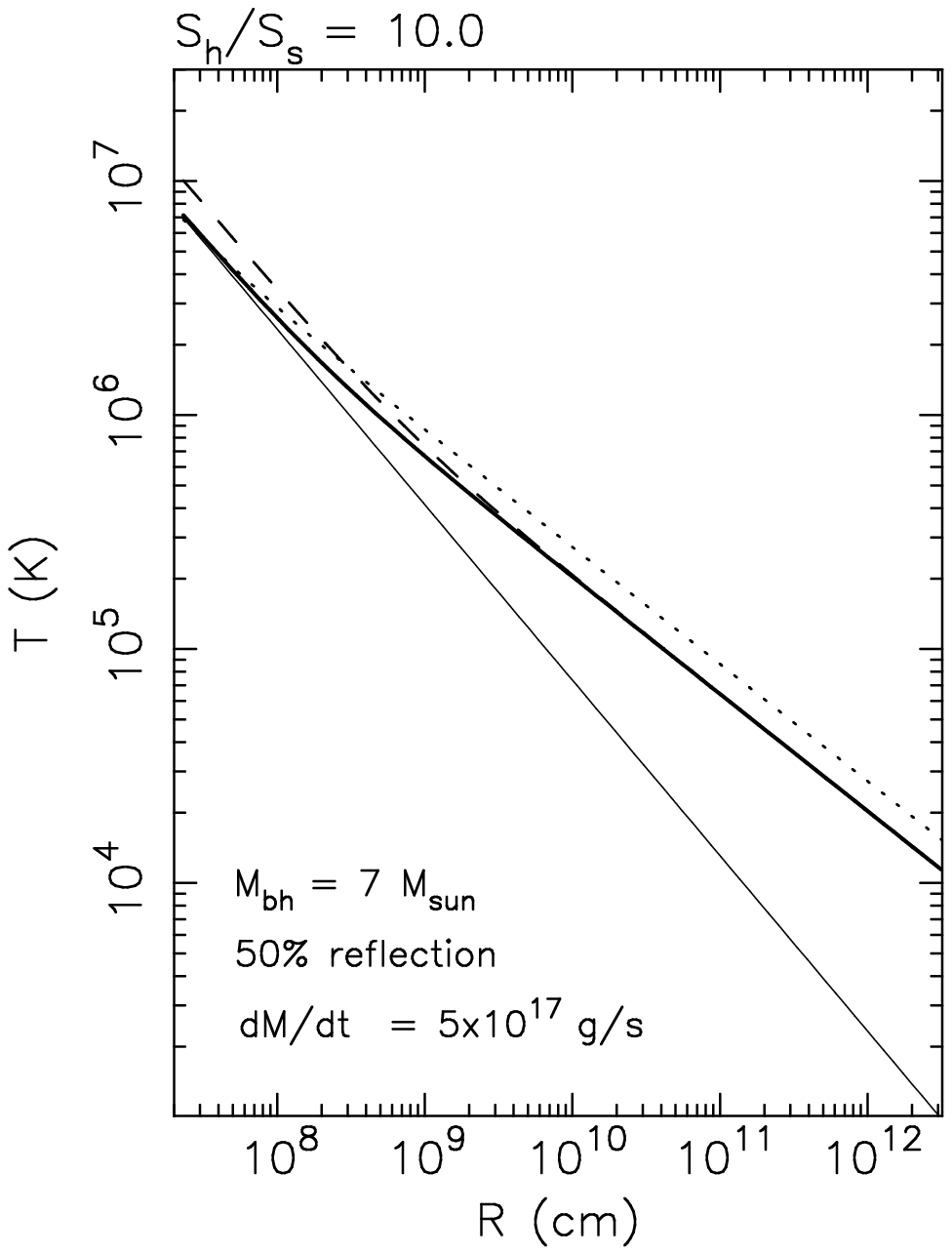} 
\end{center} 
\caption{ 
The temperature structures of an irradiatively heated accretion disk 
  calculated from the plane-parallel model. 
The surface temperature (at $\tau = 0$) of the disk is represented 
  by the dotted line; 
  the effective temperature (at $\tau = 2/3$) by the thick solid line;
  and the temperature at $\tau = 10$ by the dashed line. 
We have assumed a Shakura-Sunyaev disk 
  with a viscosity parameter $\alpha_{\rm vis} =1$. 
For comparison we also show the effective temperature (thin solid line) 
  of the same Shakura-Sunyaev disk without X-ray irradiation. 
The mass of the black hole is 7~M$_\odot$; 
  the mass-accretion rate is $5 \times 10^{17}~{\rm g~s}^{-1}$; 
  and the incident angle of the X-rays is fixed to be $87^\circ$.  
In the left panel, the disk is irradiated by X-rays 
  with a hardness ratio $\xi = 0.02$, 
  and 99\% of the X-rays are reflected.
We mark the radii at which the Balmer, He\,{\scriptsize I} 
  $\lambda$\,6678, He\,{\scriptsize I} $\lambda$\,7065, 
  He\,{\scriptsize II} $\lambda$\,4686 and 
  N\,{\scriptsize III} $\lambda \lambda$\,4641,4642 lines are emitted, 
  inferred from the peak separation       
  observed from GX~339$-$4 in 1998.   
In determining the location of the line-emission region, we assume 
  an orbital inclination $i$ of 15$^\circ$    
  and that the lines are emitted   
  from a geometrically thin, Keplerian accretion disk.  
If the period is longer or the orbital inclination is larger, 
  the line-emission regions will be at some larger radii.
H~{\scriptsize I} Balmer lines are emitted at temperatures $\sim (1 - 2) \times 10^4$~K; 
  He\,{\scriptsize I} lines at $\sim (2 - 4)  \times 10^4$~K; 
  and He\,{\scriptsize II} lines at $\sim (5 - 10) \times 10^4$~K. 
The Bowen N\,{\scriptsize III} $\lambda \lambda$\,4641,4642 lines 
  are formed in regions 
  where He\,{\scriptsize II} $\lambda$\,304 photons are abundant. 
As shown, the temperatures at the inversion layer are consistent 
  with the temperatures at which these lines are emitted. 
In the middle panel, the accretion disk is irradiated by X-rays 
  with a hardness ratio of $\xi = 50$, 
  and 99\% of the X-rays are reflected.  
In the right panel, the hardness ratio $\xi = 10$, 
  and only 50\% of the X-rays are reflected.   
\label{fig:layer}}
\end{figure}    

\section{Line morphology}  

\subsection{Asymmetric profiles and absorption troughs} 
 
The main characteristics of the optical double-peaked lines 
  of black-hole binaries 
  seen in the high-soft states are that  
  (i) the peaks are asymmetric with a stronger red peak 
  and (ii) the lines are associated with broad absorption troughs.    
As shown in the sections above, irradiative heating 
  can cause an increase in the local temperature of the accretion disk. 
If the thermal temperature of the disk surface at a radius $R$ 
  is significantly larger than the local virial temperature, 
  the disk material at the surface will evaporate.  
Irradiative heating is more effective at larger radii 
  and a wind will be formed in the outer disk regions 
  with velocities slightly higher than 
  the local virial (and Keplerian) velocities.   
Because the wind is from the outer disk,  
  the temperatures are not necessarily very high.  
The wind, which always has negative line-of-sight velocities  
  with respect to a distant observer,  
  absorbs only the blue part of the lines.   
As the wind does not absorb the red part of the line, 
  the lines appear to have a stronger red peak.   
  
The broad absorption troughs are 
  due to the absorption of the continuum by high-velocity material. 
%As the red wing of the trough is visible, 
%  the absorbers are unlikely to be in between the binary and the observer.  
If the broad absorption lines 
  originate from the accretion disk, 
  X-ray irradiative heating should be unimportant 
  in the regions where they are formed.   
A natural choice is that  
  the absorption regions are at smaller radii 
  than the regions where temperature inversion occurs. 
This is consistent with the observations that 
  the absorption troughs are broader 
  than the double-peaked emission lines 
  formed in the temperature-inversion layer. 

\subsection{Temperature-inversion vs optically thin disk} 

\begin{figure}[t]  
\begin{center} 
  \epsfxsize=16pc \epsfbox{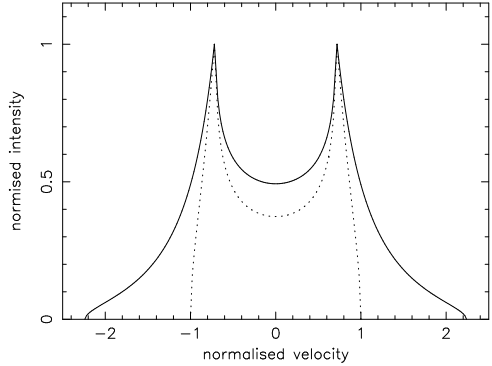}  
\end{center} 
\caption{  
Normalised profiles of double-peaked lines 
  from an optically thin, viscous heated disk (solid line) 
  and an irradiatively heated accretion disk (dotted line).  
For the viscous heated disk, 
  radiation is emitted throughout the entire viscous heated disk, 
  with a emissivity function 
  of a second-order power law of the disk radius.  
For the irradiatively heated disk,  
  only the outer disk regions, where temperature inversion occurs, 
  emit line radiation. 
Here the inner radius of the heated region 
  is assumed to be half of the accretion-disk radius, 
  and the emissivity function take the same power-law form.  
The viewing angles in both cases are 45$^\circ$.  
The line of the viscous heated disk has a standard Smak$^6$ profile, 
  but the line of the irradiatively heated disk does not.  
Note that there is a significant difference 
  in the wings of the two lines.   
} 
\end{figure}   

For an irradiatively heated accretion disk,  
  the emission lines arise  
  from the temperature-inversion layer in the outer disk region. 
As the inner disk region, where the Keplerian velocities are large, 
  does not contribute to the line emission, 
  the lines should have weak wings. 
Figure 4 illustrates the difference between the line wings 
  of an irradiatively heated disk and a conventional accretion disk, 
  for which the line assumes a Smak\cite{sxxx81} profile. 
If the emissivity of the line emission regions 
  have the same dependence on the radius, 
  the line profile of the irradiatively heated disk 
  has weaker wings and sharper peaks.  

If the accretion disk is optically thin to the optical continuum, 
  the line emission regions may cover a large portion 
  of the accretion disk, 
  which includes the inner disk region. 
As substantial emission is contributed by the high-velocity material 
  in the inner disk region, 
  the lines would have strong wings.  
As the accretion disk does not have a strong wind to absorb the line, 
  the line profiles preserve the symmetry 
  of the red and blue peaks.    
Observations\cite{jkox89} show that the H~{\scriptsize I} Balmer 
line profiles 
  of A0620$-$00 in quiescence are consistent 
  with the theoretical Smak\cite{sxxx81} profiles.   
The different morphology 
  of the doubled-peaked Balmer lines 
  observed in RXTE~J1550$-$564, GX~339$-$4 and GRO~J1655$-$40  
  during the high-soft state  
  and the lines observed in A0620$-$00 during quiescence 
  supports the interpretation that 
  emission lines are formed in a temperature-inversion layer 
  above an opaque accretion disk 
  when the system is in the high-soft state, 
  and emission lines are formed in the disk regions 
  transparent to the optical continuum when the system 
  is in an X-ray quiescent state.  

\section{Summary}     

Black-hole binaries show double-peaked emission lines   
  during the high-soft and X-ray quiescent states.   
The line profiles observed in the high-soft state 
  are asymmetric with a stronger red peak.    
Moreover, the lines are often associated with broad absorption.    
The emission lines are probably formed in a temperature-inversion layer 
  on the surface of an accretion disk optically thick 
  to the optical continuum. 
The temperature-inversion layer is due to  
  irradiative heating from regions near the accreting black holes. 
Strong temperature inversion can be formed 
  only when the X-rays have a strong soft ($\sim 0.1- 1$~keV) component. 
Irradiation by only hard ($\sim 10$~keV) X-rays 
  heats the disk uniformly at all depths, 
  and will not lead to the formation of the temperature-inversion layer. 
Irradiative heating by soft X-rays may also 
  cause the evaporation at the surface of the outer disk, 
  and hence drive a wind. 
Because of the approaching line-of-sight velocities, 
  the wind preferentially absorbs the blue part of the lines,  
  causing the asymmetry in the line profiles. 

The double-peaked lines observed in the quiescent state 
  are highly symmetric and have strong wings, 
  consistent with a theoretical Smak profile.  
As an opaque medium 
  with no temperature variations along the optical path,  
  or with a negative temperature gradient,  
  would not emit emission lines, 
  the observed double-peaked lines must be formed 
  in regions transparent to the optical continuum.   
The lines have a symmetric profile 
  because they do not encounter absorbers 
  with a specific distribution of line-of-sight velocities.

\end{document}